\documentclass[twocolumn,pra,showpacs,final,superscriptaddress,10pt,aps,floatfix]{revtex4}
\usepackage{bm}
\usepackage[usenames]{color}
\usepackage{amsmath}
\usepackage{amssymb}
\usepackage{graphicx}
\usepackage{mathrsfs}
\usepackage{subfigure}
\makeatother

\newcommand{\e}{{|e\rangle}}
\newcommand{\g}{{|g\rangle}}
\newcommand{\s}{{|s\rangle}}

\begin{document}

\title{Complete quantum state selectivity in cold molecular beams using deflection-resistant dark states in a STIRAP configuration.}

\begin{abstract}
One of the main goals of chemical dynamics is the creation of molecular beams
composed of a single
(vibrational, rotational, and magnetic) quantum state of
choice. 
In this paper we show that it is possible to achieve {\it complete} quantum state selectivity by producing resistance to electromagnetically induced deflection (EID) and that the state to be selected can be ``dialed in'' at will.
We illustrate the method by showing in detail how to purify thermal beams of the LiRb and
IF molecules to yield molecular beams composed of a variety of pre-chosen single internal quantum states
and/or superpositions of such states. We expect that this method will be implemented in all subsequent 
explorations of the fundamentals of chemical reactions and their control, and the use of cold molecules 
as a vehicle for studying some of the most profound issues of quantum dynamics.
\end{abstract}

\date{\today}
\author{Xuan Li$^1$, Asaf Eilam$^2$, and Moshe Shapiro$^{2,3}$,
~\\\it
Chemical Sciences and Ultrafast X-Ray Science Laboratory$^1$, Lawrence Berkeley National Laboratory, Berkeley, California 94720, USA
and
 Departments of Chemistry$^2$ and Physics$^3$,
The University of British Columbia
\\  2036 Main Mall, Vancouver, BC, Canada V6T 1Z1}

\maketitle
Ever since their invention, molecular beams have proven invaluable in the study of intermolecular forces, chemical reactions, interaction of matter with electromagnetic fields, scattering of atoms and molecules from surface, and structures of giant molecules~\cite{book1}.
Molecular beams have enhanced our understanding of inter- and intra-molecular dynamics, chemical reactions, interaction of light and matter, 
scattering of atoms and molecules from surfaces, and the structure of large molecules, including molecules of biological importance~\cite{Levi,zewail}. 
Usually molecular beams are composed of thermal distributions of internal states, and although their temperature can be as low as a few mK, it is clear that the ability to pre-select a {\it single} internal quantum state of choice~\cite{book} would be of tremendous
importance as it will allow for the measurement of {\it state-to-state} differential cross-sections, the most detailed entity that can be 
observed in the context of molecular collisions~\cite{chemrev}.
The pre-selection of individual rotational states is also a good starting points for the clean preparation of aligned~\cite{aliment}
and oriented~\cite{orientation} molecules. 
Moreover, since molecular beams are nearly decoherence free, the formation of molecular beams containing ``tailor made'' {\it superpositions} of internal states would be highly desirable for quantum encoding and quantum memory storage.

In the past a number of directions of purifying thermal 
molecular beams have been
pursued. These techniques include: hexapole devices 
to single out one out of two rotational states~\cite{hexapole1} or the lowest 
(or highest) of many~states~\cite{hexapole2} and the deflection
of ultracold molecules via the use of optically induced dipole forces 
in far off-resonance traps (FORT)~\cite{FORT}. 
While the FORT method may be an effective method of preparing molecules 
in a chosen rotational state, the method cannot induce vibrational
selectivity. Thus,
 the general task of selectively preparing molecular beams
composed of a {\it single} ro-vibrational state of our choice 
has so far not been achieved.

In this paper we show how, using Coherent Control 
techniques~\cite{book}, one can convert thermal molecular beams 
into beams containing only a {\it single} pre-selected
(ro-vibrational and magnetic) internal state.  The method, based on the
electromagnetic preparation
of ``deflection-resistant'' states, is best explained via a model 
 composed of three molecular states 
($\g$, $\e$, $\s$)  
interacting with two spatially inhomogeneous 
electromagnetic fields, 
the ``probe'' field $\varepsilon_p(X)$ and 
the ``control'' field $\varepsilon_c(X),$ where $X$ is the
spatial variable. The $\varepsilon_p(X)$ and $\varepsilon_c(X)$ fields,
whose effective strengths are parameterized by the Rabi frequencies 
$\Omega_p(X)=\varepsilon_p(X)\cdot\mu_{e,g}/2\hbar$ and 
$\Omega_c(X)=\varepsilon_c(X)\cdot\mu_{s,e}/2\hbar,$ 
where 
$\mu_{e,g}$ and $\mu_{s,e}$ are
the transition dipoles for the 
$\g\leftrightarrow \e$ and $\e\leftrightarrow \s$ transitions,
are energetically detuned from these one-photon 
transitions by $\delta_p$ and $\delta_c,$ respectively. 

\begin{figure}
\centering
\includegraphics[width=0.8\columnwidth,angle=0]{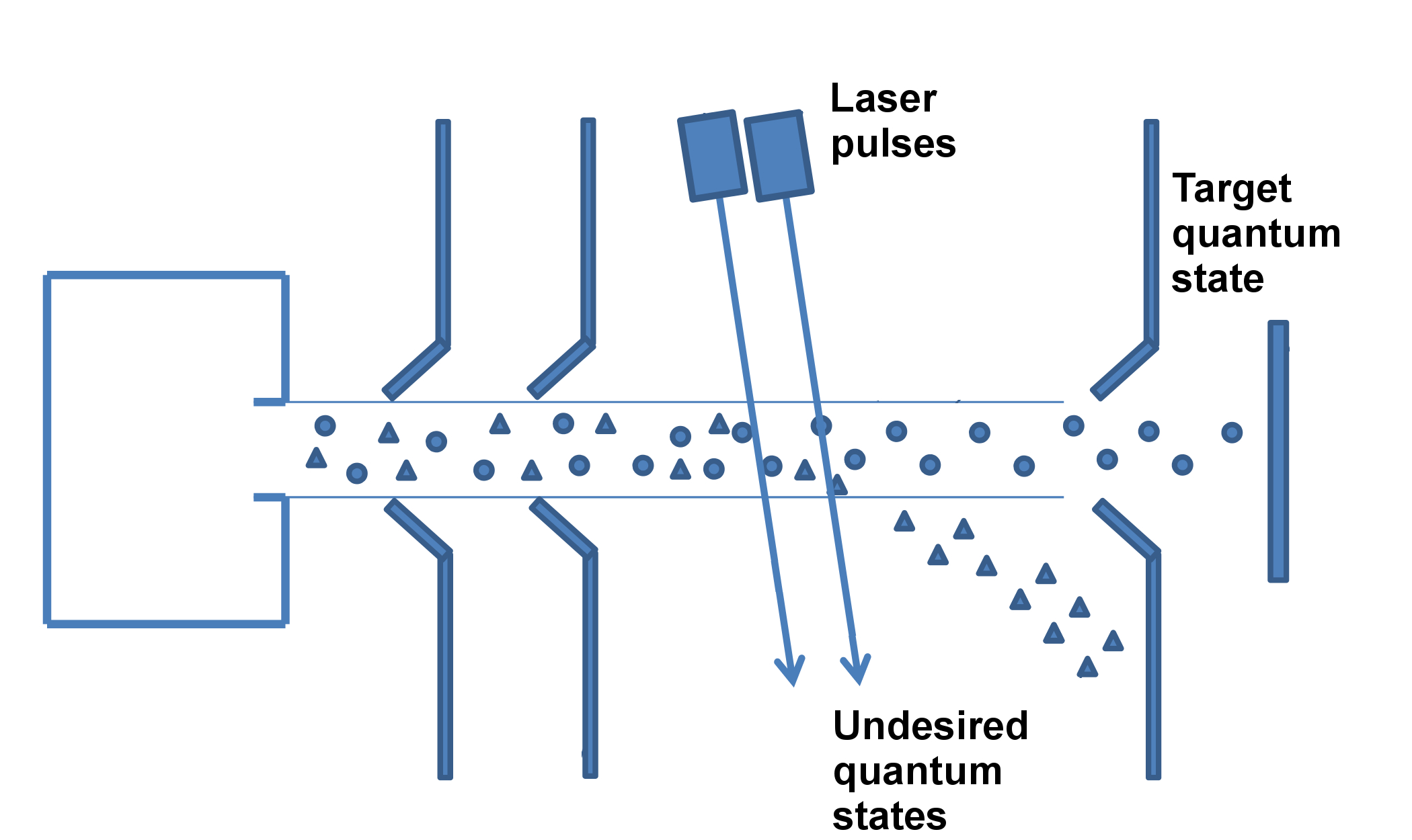}
\vskip .5truein
\caption{{\small(Color online) The purification scheme: 
target state molecules (filled circles) sail through the three slits; 
molecules in other states (filled triangles) get deflected. Note, the laser beam is focused slightly off
axis relative to the molecular beam to maximize the energetic gradient.}} \label{scheme}
\end{figure}

Relative to the FORT~\cite{FORT} method, our method {\it reverses} 
the roles of the state to be selected and the states to be
discarded: The discarded states are those that suffer deflection, while the
selected state is an electromagnetically generated ``deflection-resistant'' 
dark state. 
Molecules in the discarded states simply experience an effective dipole force due to the lack of the
two-photon resonance. 
At the same time we render our chosen (vib-rotational) state $\g,$  
``deflection resistant'' by tuning the two fields 
to be in {\it two-photon} resonance with
the $\g\leftrightarrow\e\leftrightarrow\s$ transition. 
The chosen state become transparent to the probe field 
because it evolves into a {\it dark state} 
of the same type encountered in {\it ``Coherent Population 
Trapping''} (CPT)~\cite{Arimondo1976,Arimondo1996}, {\it ``Stimulated Raman Adiabatic Passage''} (STIRAP)~\cite{Bergmann}, and 
{\it ``Electromagnetically Induced Transparency''} (EIT)
~\cite{Boller1991,Field1991,Harris1997,fleishhauer05}. 
Such dark states are known to result from the
a destructive interference between two overlapping resonances arising from two
Autler-Townes~\cite{Autler1955} split dressed-states that are broadened due to 
the presence of a continuum~\cite{Shapiro2007,Eilam12} 
(e.g. that of the spontaneously emitted photons from the $\e$ state). 
As a result of the interference, a transparency
window in which no absorption of the probe field occurs
is formed at $\omega_{e,g}=(E_e-E_g)/\hbar$, 
justifying the name ``dark state''.  
Since the dark state is oblivious to the
field, it feels no force whatsoever and sails 
through the three
slits of the setup depicted in Fig. \ref{scheme} in a straight line, 
ending up as the only state left in the beam. An approach for deflection of
polaritons induced by an EIT dark state ~\cite{Karpa2006}  differs from the current proposal in 
the origin of the forces, experimental set-up, and temperature requirement.

We now detail the use of this method for quantum state selectivity: 
As shown in \ref{scheme}, a supersonic molecular beam collimated by 
two sequential skimmers is irradiated by a spatially inhomogeneous 
control field ($\varepsilon_{c}(X)$) 
somewhat detuned from the 
$\omega_{e,s}$ transition frequency. 
Further down-stream a Gaussian shaped spatially inhomogeneous probe field 
($\varepsilon_p(X)$)
irradiates the molecular beam. Its frequency is chosen such that the 
two-photon resonance condition, $\delta_{two}=\delta_p-\delta_c=0,$
is satisfied. 
By positioning the laser beams such that the 
molecules are subjected to $\varepsilon_c(X)$ 
{\it before} they see $\varepsilon_p(X)$ 
(the ``counter intuitive ordering''\cite{Bergmann,Bergmann2}), 
we make sure that the $\g$ state we wish to select will in fact evolve 
into a dark state.

The dots of \ref{scheme}  represent dark state 
molecules that are not subject to any force and sail through the slits in a straight line. The triangles of \ref{scheme} represent 
molecules which are in other internal states and thus do not satisfy the 
two-photon resonance and are subject to a non-zero force. Such molecules
are deflected away.  
We will give a possible set-up as an example. For a $10$ micron laser focus and a $1$ degree angle between 
the laser beam and the molecular beam, the effective length of the laser profile seen by
the molecular beam is $\sim 600$ microns; in addition, the Rayleigh length of a $586$ nm laser, as used in our numerical simulation,
is estimated to be $750$ microns given a $10$ cm focal length and $6.3$ mm aperture diameter. Both of these lengths determines
the time scale of the interaction time: a few $\mu$s. 

In order to demonstrate that the system follows adiabatically
a particular field-dressed state, we first obtain 
the system wave function by expanding it in the bare states basis  
$|\Psi\rangle=\g\Psi_g+\e\Psi_e+\s\Psi_s,$ and solve 
for the expansion coefficients $\Psi_{g/e/s}(t)$ which satisfy the matrix time-dependent 
Schr\"odinger equation 
\begin{equation}
\label{schroed}
i\frac{\partial}{\partial t}
\left|
\Psi(X,t) 
\right\rangle
=
\underline{\underline{\bf{H}}}(X,t) 
\cdot
\left|
\Psi(X,t) 
\right\rangle
\end{equation}
where
\begin{eqnarray}
\underline{\underline{\bf H}}(X,t)
&=&2\left[\Omega_p(X,t)\left|g\right\rangle \left\langle e\right|+\Omega_c(X,t)\left|s\right\rangle \left\langle e\right|\right.\nonumber\\
&+&{\delta_p\over 2}\left|e\right\rangle \left\langle e\right|+ \left.{\delta_{two}\over 2}\left|s\right\rangle \left\langle s\right|+c.c.\right]
\end{eqnarray}
with $\Omega_p={\bf \mu}_{e,g}\cdot{\bf \varepsilon}_p/2\hbar$ and
$\Omega_c={\bf \mu}_{s,e}\cdot{\bf \varepsilon}_c/2\hbar$.  
In addition, for both $\delta_{two}=0$ and $\delta_{two}\not=0$, we find the three adiabatic states $\underline{\psi}^{(1,2,3)}$ and their eigenenergies ${\cal E}^{(1,2,3)}$.
For the target state, $\delta_{two}=0$ is chosen and one of the eigenenergies is ${\cal E}^{(1)}=0,$ i.e., a {\it dark state}. 
When the molecule encounters $\varepsilon_c(X)$ spatially before $\varepsilon_p(X)$
(the ``counter-intuitive ordering''\cite{Bergmann}), $\underline{\psi}^{(1)}$,
corresponding to ${\cal E}^{(1)}=0,$ is the only time-dependent field-dressed state that correlates
with the chosen bare state $\g,$ at $t=-\infty$ which, therefore, continues to
evolve as a dark state, oblivious to the external fields. For other states,
where $\delta_{two}\not=0$, the system adiabaticly follows a particular field-dressed state, the field-induced potential of which is not homogenous in space. In all of our numerical simulations, the projection of 
this adiabatic state to the ``trapping state'' in FORT~\cite{FORT} is extremely close to one. Therefore, these other states, where $\delta_{two}\not=0$,
are subject to deflection forces.

\vskip 0.2 truein
\begin{figure}[ht!]
                \centering
              \includegraphics[width=0.7\columnwidth,clip=true]{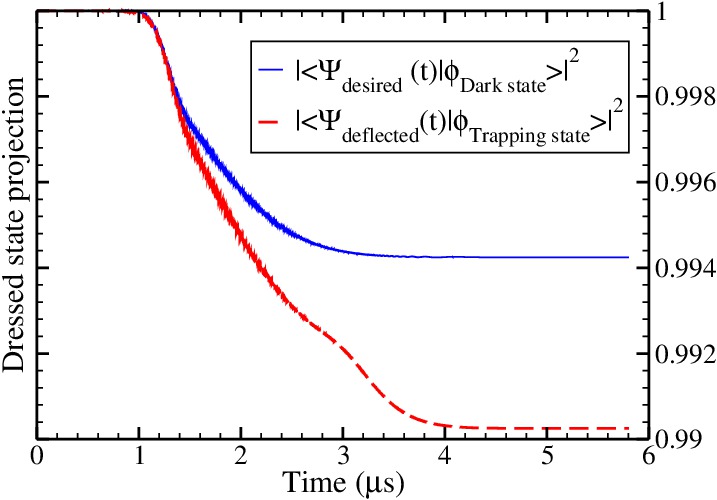}
\caption{{\small Adiabatic following of the desired state (LiRb, $\nu_X=0, J_X=0$) in solid blue and
 the deflected state (LiRb, $\nu_X=0, J_X=1$) in dashed red. }
} \label{adiabatic}
\end{figure}

This much simplified three-level model we described in \ref{schroed} is adopted to give a straight-forward description of adiabatic following of different field-dressed state, a field-resistant dark state or the deflected trapping state, of the molecules depending on $\delta_{two}$. However, to be more realistic, there are three types of terms that may modify the Hamiltonian in \ref{schroed} to disturb or even destroy the adiabatic following due to a finite one-photon detuning and intense laser fields applied:  {\it counter-rotating terms}, coupling to {\it other intermediate states}, and other field dressing terms due to pulse mixing where the ``probe'' field can act as ``control'' field, vice versa. The details of these additional terms will be given in the supplemental material and we will only summarize the conclusion here. 
In order to test whether our previous conclusion by using a simplified three-level model still holds in the presence of these additional terms, and to evaluate the effective optical potential imposed on a specific state, we incorporate all of these additional terms in our numerical simulations and expand the three-level models to 7-,9- and 11-level models for convergence studies. 
In the many-level simulations, we project the simulated time-dependent wave packet of the molecules onto the field-dressed states. As shown in  \ref{adiabatic}, we observe that nearly all the molecules are following the field-dressed states, i.e. the desired molecules are adiabatically the field-resistant dark state and the deflected molecules are following the trapping state. The observed nearly perfect adiabatic following proves the validity of the optical potentials, which are later applied to the 
 classical motion study,  and that these optical potentials correspond exactly to their equivalents in a simple three level model.  
 Note, due to inevitable scattering of the molecules with the photons and the subsequent spontaneous emission loss,
 the projection of the total wave function onto the field dressed states is close to but not equal to unity.  

By solving the 
full time dependent Schr\"odinger equation for the multi-level models,
and ascertaining that the molecular system indeed follows 
one of the adiabatic states , we can then confidently obtain 
the classical forces acting on the 
molecules, given as $F(X,t)=
-{\partial\over \partial X} {\cal E}(X,t)$ and 
compute the classical trajectories of molecules in 
different quantum states subject to these forces. 

As a first example we consider the purification of a thermal 
beam of LiRb molecules. In the arrangement depicted in 
Fig. \ref{scheme}, the average longitudinal 
velocity of the LiRb beam as it passes through the skimmers 
is $\sim 500$ m/s. The sequential-slits arrangement forces the average 
transverse velocity to be zero, while the  
supersonic expansion reduces the longitudinal and transverse 
temperatures to values as low as $\sim 5^o$K. Since the
vibrational and rotational temperatures follow the translational temperature, 
only the ground ($\nu_X=0$) 
LiRb vibrational level, 
where $X$ denotes the $X^{1}\Sigma$ ground electronic state, 
is populated at $ 5^o$K,
with the higher vibrational states 
remaining empty. In contrast, the range of the rotational 
levels that are populated at this temperature 
is $J_X=0$ to $J_X=7$, with a rotational constant $B_e=0.2158$ cm$^{-1}$. 

\vskip 0.2 truein
\begin{figure}[ht!]
               \includegraphics[width=0.6\columnwidth,clip=true,angle=-90]{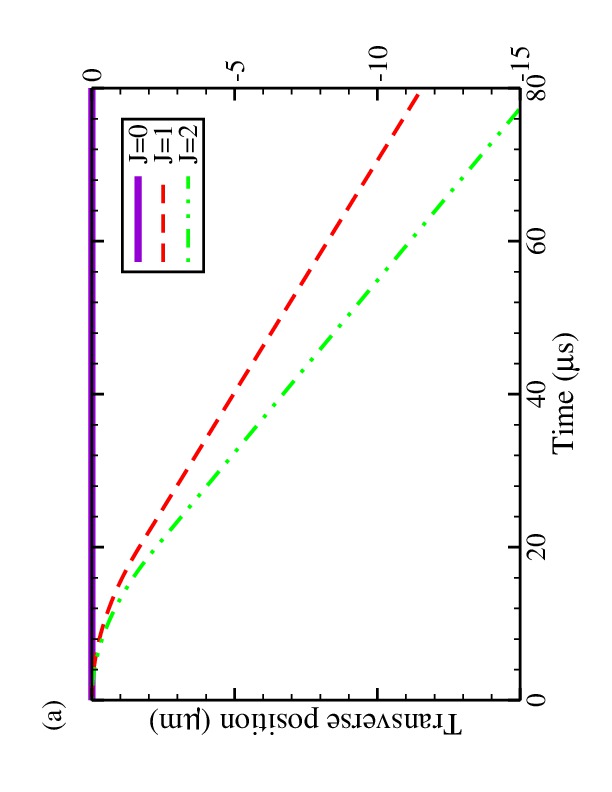}
              \includegraphics[width=0.6\columnwidth,clip=true,angle=-90]{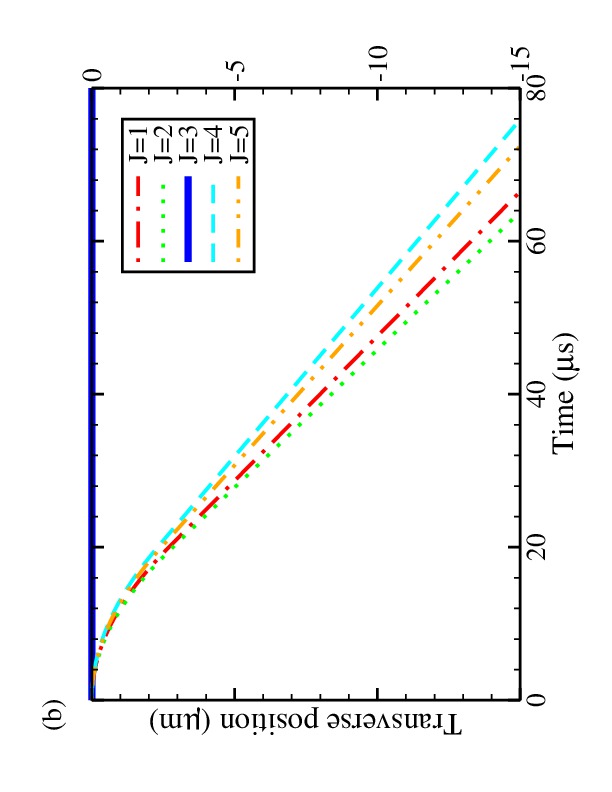}
\caption{{\small Molecular beam purification into different rotational states. 
Shown is the spatial evolution in the transverse direction as a function of 
time for: {\bf (a)} the 
$\left|\nu=0,J=0\right\rangle$ target quantum state (purple solid line); 
{\bf (b)} the $\left|\nu=0,J=3\right\rangle$ 
target quantum state (blue solid line).
}
} \label{figure3}
\end{figure}

\begin{figure}
\centering
\includegraphics[width=0.5\columnwidth,clip=true,angle=-90]{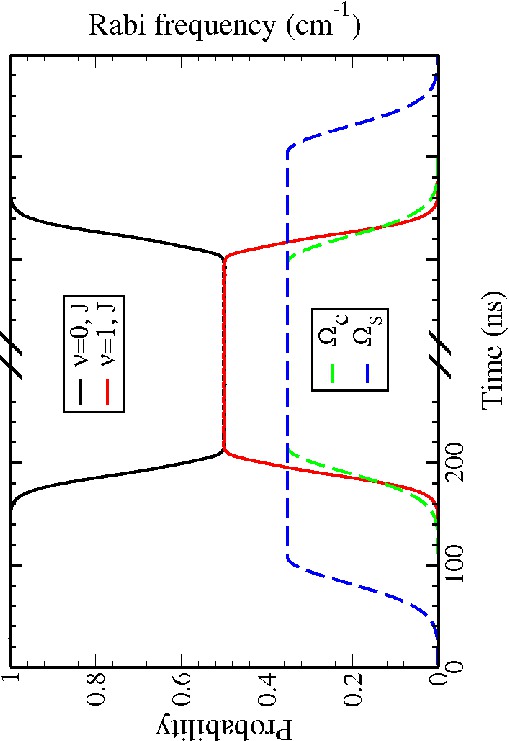}
\caption{{\small 
(Color online) Making just one bare state ($|\nu=0\rangle$)
deflection resistant by forming and unforming a 
transient superposition with the $|\nu=1\rangle$ state.
Shown is the time dependence of the 
populations of the two states 
and that of the laser fields. 
The turn-off of the fields is the time reversal of their 
turn-on.}} \label{figure4}
\end{figure}

To demonstrate
the ability to purify the system into {\it any} pre-selected rotational state, 
we study two cases in which we create a beam composed entirely of the
$\left|\nu_{X}=0,J_{X}=0\right\rangle$ state or the 
$\left|\nu_{X}=0,J_{X}=3\right\rangle$ state.
Numerical simulations of classical trajectories under the action of the
optical fields when 
$\left|\nu_{X}=0,J_{X}=0\right\rangle$ or $\left|\nu_{X}=0,
J_{X}=3\right\rangle$ is selected, 
are presented in \ref{figure3}a and \ref{figure3}b, 
respectively. Both the 
$\varepsilon_p$ and $\varepsilon_c$ electromagnetic fields, 
have powers of $0.8$ W and waist sizes of $10$ $\mu m$. 
The one-photon detuning of the probe field, $\delta_{p}$, is set to 
be $300$ cm$^{-1}$ on the red side of 
$\left|\nu_{B}=0,J_{B}=J-1\right\rangle$. 
As shown in \ref{figure3}a (b), the target states 
($\left|\nu=0,J=0\right\rangle$ or $\left|\nu=0,J=3\right\rangle$), 
are made to evolve into field-dressed dark states 
by setting the laser parameters 
to fulfill the two-photon resonance condition $\delta_{two}=0.$ In contrast, 
all other states, $\left|\nu=0,J\neq0\right\rangle$
($\left|\nu=0,J\neq3\right\rangle$), do not satisfy this condition and
molecules in these undesired states 
are thus deflected towards the high-field region of the laser waist and 
move away from the molecular beam. 
Furthermore, the trajectories of the deflected molecules in various quantum states are slightly different, implying different forces due to different
$\delta_{two}$. Regarding the temporal evolution of the wave function of the
target molecules, it is well established that keeping both lasers on will
conserve the quantum superposition between the two ground states
($\g=\left|\nu=0, J\right\rangle$ and $\s=\left|\nu=1,
J\right\rangle$). However, if we adiabatically turn off the probe field before
the control field, after the separation is complete, 
we can transfer the population back to the single target state, $\left|\nu=0,
J\right\rangle$. 
The temporal sequencing of the electromagnetic fields during the whole process, on and off of the fields, is presented on \ref{figure4}, where we can see the adiabatic evolution of the population in states $\g$ and $\s$. 

The thermal spreading should also be considered to demonstrate the applicability of such a separation scheme in reality. For {\it longitudinal thermal spreading} the main effect is Doppler broadening. For a beam with longitudinal central
velocity of $500$ m/s, $\delta_v$ - the longitudinal velocity spread is on the order of $10\%$ i.e.,  $\sim50$
m/s. This results in relative frequency spread of $\delta_\omega/ \omega=1.6 \times10^{-7}$ which
is negligible relative to both the one-photon detuning and the detuning from two-photon resonance. In the {\it transverse direction}, for the {\it deflected molecules}
we discard, a strong enough light field guarantees
that the light induced deflection potential these molecules experience 
is greater than the transverse thermal spread in kinetic energy.  
Our experience is that as long as 
$\Omega^2_{max}/\delta > 5\cdot m v_{t}^2/2$, where $v_{t}$ is the velocity 
in the transverse direction,  the separation scheme works extremely well. 
$v_t$ can be controlled to some extent by tuning the ratio between 
the slit size and the longitudinal distance between the first two slits.  



The results presented here display a high degree of selectivity with respect to
the two photon resonance condition. For example, if we set $\delta_{two}=0$
for a $\Lambda$ system originating in the LiRb 
$\left|\nu_X=0,J_X=0\right\rangle$ state, 
$\delta_{two}$ for the $\left|\nu_X=0,J_X=1\right\rangle$ state 
will be as small as 
$0.003$ cm$^{-1}=90$ MHz. Yet the discrimination between these two states
works well! Such great resolution makes it possible to 
differentiate between hyperfine levels, $f$, or even various $m_f$
sub-levels. For example, in atomic $^6$Li 
subject to moderate magnetic fields, the splitting between 
the $\left|f=1/2,m_f=1/2\right\rangle$ level and the  
$\left|f=1/2,m_f=-1/2\right\rangle$ level is $75$ MHz, 
enough to allow for the selective preparation of a
beam composed of only one of these states by the present method.

One of the merits of such a proposal lies in the robustness against realistic imperfect laser conditions. Such an advantage over methods that {\it collect} the deflected molecules is
that small effects on the trajectories of the deflected molecules are irrelevant because on the {\it non-deflected} molecules are collected, as long as the deflection is large enough to prohibit
the arrival of the deflected molecules to the third slit, as shown in \ref{scheme}. As for the {\it non-deflected} molecules, their properties are due to the formation of dark states, which are the heart
of the STIRAP and EIT experiments, and that it was shown in the past, both theoretically and experimentally, that these processes are robust against laser imperfections such as, intensity fluctuations, frequency drifts, and imperfect laser beam overlaps~\cite{Bergmann}. More detailed discussions on the robustness can be found in the supplementary material (section E).

The most immediate future application of our method is the selective preparation of 
beams containing {\it superposition  states,} such as those existing in 
 \ref{figure4} when the two pulses are on. 
Such ``superposition beams,'' whose coherence is expected to survive
for long times because the decoherence processes inside the
beam are extremely slow, would find use in 
quantum memory and quantum computation devices.
Another future direction would involve the use of 
broad-band, short laser pulses. In
the straightforward application of the present technique, the molecular 
level spacings set a lower limit on the pulse bandwidth to be used, hence on 
the time scale of the separation process. In order to speed up 
the process by employing broadband lasers of short 
($\sim 1$ ps) durations and enjoy the higher field strengths associated with
such lasers, we propose to use Coherent Control pulse-shaping 
techniques\cite{silberberg01a} which allow one to 
home in on a desired two photon transition and render the state of
our choice ``deflection-resistant'' even when the broad
laser spectrum covers a number of transitions which satisfy the two photon
resonance condition.   
Another future application would involve the use of  different polarizations, 
or spatially inhomogeneous magnetic fields to deflect away unwanted $m_j$ states. 
We can also envision the discrimination between 
states belonging to multiple minima in big molecules~\cite{Meerakker}
and deflecting out minority molecules in a mixture in the 
presence of majority molecules that absorb in the same spectral 
region~\cite{Eilam12a}.

\end{document}